# Interference Alignment via Improved Subspace Conditioning

Douglas Kim and Murat Torlak
University of Texas at Dallas
Department of Electrical Engineering
Richardson, TX 75080, USA
{dekim, torlak}@utdallas.edu

*Abstract*—For the $K$ user, single input single output (SISO), frequency selective interference channel, a new low complexity transmit beamforming design that improves the achievable sum rate is presented. Jointly employing the interference alignment (IA) scheme presented by Cadambe and Jafar in [1] and linear minimum mean square error (MMSE) decoding at the transmitters and receivers, respectively, the new IA precoding design improves the average sum rate while preserving the achievable degrees of freedom of the Cadambe and Jafar scheme, $K/2$.

## I. INTRODUCTION

Cadambe and Jafar (CJ) have shown in [1] that for the $K$ user, single input single output (SISO), frequency selective interference channel, $K/2$ degrees of freedom is achievable by using interference alignment (IA) at the transmitters and zero forcing (ZF) decoding at the receivers. Recent work by Shen, Host-Madsen, and Vidal (SHV) has shown that the joint IA and ZF strategy can be further improved to increase the data rate performance of the overall system while maintaining the achievable degrees of freedom in the following two ways. One in which new precoding subspaces are used to generate the precoding vectors and another which optimizes the precoding vectors within the subspaces defined in the original CJ scheme. While the optimization of the second method is presented in [2], an explicit solution to the first is not.

Then in [3], Kim and Torlak (KT) proved that the sum rate of the $K = 3$ SISO interference alignment channel in which receivers employ linear minimum mean square error (MMSE) decoders is a concave function with respect to $\boldsymbol{w}$, (12) in Sec. II, a vector with which the precoding subspaces are defined. Moreover, they prove that the concave sum rate expression can be formulated into a constrained maximization problem for which a global solution exists thereby improving the conditioning of the precoding subspaces and maximizing the sum rate with respect to $\boldsymbol{w}$.

However, calculation of the optimal solution requires use of a first order optimization algorithm, i.e. gradient descent. Therefore, as a trade off in data rate performance for reduced computational complexity, KT also derive a lower bound to the sum rate expression that when maximized yields a solution that is far less complex to calculate but still achieves significant increases in sum rate while also maintaining the achievable degrees of freedom of the CJ scheme.

In this paper, a new lower bound to the sum rate expression introduced in [3] is derived that when maximized produces a simple *closed-form* solution that greatly reduces computational costs compared to either of the solutions proposed by KT while still providing considerable gains in data rate performance. Moreover, the new closed-form solution also maintains the achievable degrees of freedom proposed by CJ.

Notations: we use upper case letters to denote matrices, lower case letters to denote scalars, and boldface letters to denote vectors. $(\cdot)^t$ and $(\cdot)^\dagger$ refers to the transpose and conjugate transpose of $(\cdot)$, respectively.

## II. SYSTEM MODEL

Consider the $K = 3$ user interference channel where user 1 is allocated $n + 1$ streams of data while users 2 and 3 are both allocated $n$; and all three users transmit over a $2n + 1$ symbol extension of the channel by means of beamforming. Here, $n \in \mathbb{N}$. The channel output at the $k$th receiver is then defined as

$$\boldsymbol{y}_k := H_{k1}V_1\boldsymbol{x}_1 + H_{k2}V_2\boldsymbol{x}_2 + H_{k3}V_3\boldsymbol{x}_3 + \boldsymbol{z}_k$$

where $k = \{1, 2, 3\}$ is the user index. $\boldsymbol{y}_k$ is an $N$ by 1 vector representing the output signal of the $k$th receiver where $N = 2n + 1$. $H_{kj}$ is a diagonal $N$ by $N$ matrix of channel fading coefficients from transmitter $j$ to receiver $k$.

The $(n + 1)$ by 1 vector $\boldsymbol{x}_1$ represents the input signal of the first transmitter and $\boldsymbol{x}_k$ for $k = \{2, 3\}$ is an $n$ by 1 vector representing the input signal of the $k$th transmitter. Vector $\boldsymbol{x}_1 \sim \mathcal{CN}(0, pI_{n+1})$, and $\boldsymbol{x}_2$, $\boldsymbol{x}_3$ are both $\mathcal{CN}(0, pI_n)$.

Precoding matrix $V_1$ is $N$ by $(n + 1)$ and matrices $V_2$ and $V_3$ are both $N$ by $n$.

Lastly, $\boldsymbol{z}_k$ is an $N$ by 1 vector representing the additive white Gaussian noise (AWGN), i.e. $\boldsymbol{z}_k \sim \mathcal{CN}(0, I_N)$.

In order to guarantee the alignment of interference at each of the three receivers, CJ define the matrices of beamforming vectors as

$$V_1 := A \qquad (1)$$
$$V_2 := (H_{32})^{-1}H_{31}C \qquad (2)$$
$$V_3 := (H_{23})^{-1}H_{21}B \qquad (3)$$

where

$$T := H_{12}(H_{21})^{-1}H_{23}(H_{32})^{-1}H_{31}(H_{13})^{-1} \quad (4)$$
$$A := \begin{bmatrix} \boldsymbol{w} & T\boldsymbol{w} & T^2\boldsymbol{w} & \ldots & T^n\boldsymbol{w} \end{bmatrix} \quad (5)$$
$$B := \begin{bmatrix} T\boldsymbol{w} & T^2\boldsymbol{w} & \ldots & T^n\boldsymbol{w} \end{bmatrix} \quad (6)$$
$$C := \begin{bmatrix} \boldsymbol{w} & T\boldsymbol{w} & \ldots & T^{n-1}\boldsymbol{w} \end{bmatrix} \quad (7)$$

and the $N$ by 1 vector $\boldsymbol{w}$ for the CJ scheme is a vector of all ones.

Once aligned, the interference at receivers 1, 2, and 3 are constrained respectively to the following.

$$H_{12}V_2 = H_{13}V_3 \quad (8)$$
$$H_{23}V_3 \prec H_{21}V_1 \quad (9)$$
$$H_{32}V_2 \prec H_{31}V_1 \quad (10)$$

where $A \prec B$ denotes the set of column vectors of matrix $A$ is a subset of the column vectors of matrix $B$. More precisely, the set of columns in $H_{23}V_3$ are equal to the *last* set of $n$ columns in matrix $H_{21}V_1$. Likewise, the set of columns in $H_{32}V_2$ are equal to the *first* set of $n$ columns in matrix $H_{31}V_1$. For example, if $H_{32}V_2 = H_{31}C$, then

$$H_{31}V_1 = H_{31}\begin{bmatrix} C & T^n\boldsymbol{w} \end{bmatrix} = \begin{bmatrix} H_{32}V_2 & H_{31}T^n\boldsymbol{w} \end{bmatrix}.$$

In [3], the three beamforming matrices are redefined as

$$V_k(\boldsymbol{w}) := W(\boldsymbol{w})\Gamma_k \quad (11)$$

where the $N$ by $N$ diagonal matrix, $W(\boldsymbol{w})$, is defined as a function of the $N$ by 1 vector

$$\boldsymbol{w} = \begin{bmatrix} w_1 & w_2 & \ldots & w_N \end{bmatrix}^t \quad (12)$$

such that $(W(\boldsymbol{w}))_{ii} := w_i$ and $w_i \in \mathbb{R}^+$ for all $i$. Note that by setting $\boldsymbol{w}$ equal to $\mathbf{1}$, the new expression for the beamforming matrices in (11) is equivalent to the set of precoders for the CJ scheme. But by factoring out the diagonal elements of $W$ from $V_k$ and defining them as variable, a new $\boldsymbol{w}$ can be found that improves the sum rate performance.

The $N$ by $n+1$ matrix $\Gamma_1$ is defined as

$$\Gamma_1 := \begin{bmatrix} \mathbf{1} & \boldsymbol{t}_1 & \cdots & \boldsymbol{t}_n \end{bmatrix}$$

and the $N$ by $n$ matrices $\Gamma_2$ and $\Gamma_3$ are respectively defined as

$$\Gamma_2 := (H_{32})^{-1}H_{31}\begin{bmatrix} \mathbf{1} & \boldsymbol{t}_1 & \cdots & \boldsymbol{t}_{n-1} \end{bmatrix}$$
$$\Gamma_3 := (H_{23})^{-1}H_{21}\begin{bmatrix} \boldsymbol{t}_1 & \boldsymbol{t}_2 & \cdots & \boldsymbol{t}_n \end{bmatrix}$$

where $\boldsymbol{t}_m = \text{diag}(T^m)$ for $m = \{1,...,n\}$. The function $\text{diag}(\cdot)$, like that within Matlab, creates a column vector comprised of the diagonal elements of its matrix input.

## III. ACHIEVABLE SUM RATE AND LOWER BOUND APPROXIMATION

In this section, we introduce the sum rate expression derived in [3] with which we formulate a new lower bound.

### A. Achievable Sum Rate

Introduced in [3], the sum rate for the $K = 3$ SISO interference channel in which receivers employ linear MMSE decoders is defined as

$$f(\boldsymbol{w}) := \frac{1}{N}\sum_k f_k(\boldsymbol{w}) \quad (13)$$

where the individual rate for user $k$ is defined as

$$f_k(\boldsymbol{w}) := \log \frac{\det\left(I + p\sum_j H_{kj}W\Gamma_j\Gamma_j^\dagger W^\dagger H_{kj}^\dagger\right)}{\det\left(I + p\sum_{j\neq k} H_{kj}W\Gamma_j\Gamma_j^\dagger W^\dagger H_{kj}^\dagger\right)} \quad (14)$$

for $j = \{1,2,3\}$. Note that the dependence on vector $\boldsymbol{w}$ in (14) is implied to be within diagonal matrix $W$ as in (11) though not explicitly expressed here nor for the remainder of this paper as for the sake of brevity.

In Lemma 1 of [3], equation (13) is expressed equivalently in the following compact form

$$f(\tilde{\boldsymbol{w}}) = \frac{1}{N}\sum_k \Big(\log\det\left(I + pG_{kn}^\dagger \tilde{W} G_{kn}\right) - \log\det\left(I + pG_{kd}^\dagger \tilde{W} G_{kd}\right)\Big) \quad (15)$$

where $\tilde{W} := W^\dagger W$ and $\tilde{\boldsymbol{w}}$ is the vector formed of the $N$ diagonal elements of $\tilde{W}$, i.e. $\tilde{w}_i := |w_i|^2$. Furthermore,

$$G_{1n} := \begin{bmatrix} H_{11}\Gamma_1 & G_{1d} \end{bmatrix} \quad (16)$$
$$G_{1d} := \sqrt{2}H_{12}\Gamma_2 \quad (17)$$
$$G_{2n} := \begin{bmatrix} H_{22}\Gamma_2 & G_{2d} \end{bmatrix} \quad (18)$$
$$G_{2d} := H_{21}\Gamma_1 P_2 \quad (19)$$
$$G_{3n} := \begin{bmatrix} H_{33}\Gamma_3 & G_{3d} \end{bmatrix} \quad (20)$$
$$G_{3d} := H_{31}\Gamma_1 P_3. \quad (21)$$

and

$$P_2 := \begin{bmatrix} 1 & \mathbf{0}_{1\times n} \\ \mathbf{0}_{n\times 1} & \sqrt{2}I_n \end{bmatrix}$$
$$P_3 := \begin{bmatrix} \sqrt{2}I_n & \mathbf{0}_{n\times 1} \\ \mathbf{0}_{1\times n} & 1 \end{bmatrix}.$$

It is based upon this expression of the sum rate which we derive our new lower bound.

### B. Achievable Sum Rate Lower Bound

In deriving the lower bound to (15), we begin by taking its limit as $p$ goes to infinity as shown in (22) at the top of the next page.

Next, we derive an upper bound to the magnitude of the final negative term in (22), denoted as $f_{ub}(\tilde{\boldsymbol{w}})$, that in turn

$$\lim_{p\to\infty} f(\tilde{\boldsymbol{w}}) = \frac{1}{N}\sum_k \log\det\left(pG_{kn}^\dagger \tilde{W} G_{kn}\right) - \frac{1}{N}\sum_k \log\det\left(pG_{kd}^\dagger \tilde{W} G_{kd}\right)$$
$$= \frac{3(2n+1)}{N}\log p + \frac{1}{N}\sum_k \log\det\left(G_{kn}^\dagger \tilde{W} G_{kn}\right) - \frac{(3n+2)}{N}\log p - \frac{1}{N}\sum_k \log\det\left(G_{kd}^\dagger \tilde{W} G_{kd}\right)$$
$$= \frac{3n+1}{2n+1}\log p + \frac{3}{N}\log\det \tilde{W} + \frac{1}{N}\sum_k \log\det\left(G_{kn}G_{kn}^\dagger\right) - \frac{1}{N}\sum_k \log\det\left(G_{kd}^\dagger \tilde{W} G_{kd}\right) \quad (22)$$

---

lower bounds the high SNR approximation of the sum rate.

$$f_{ub}(\tilde{\boldsymbol{w}}) = \left|-\frac{1}{N}\sum_k \log\det\left(G_{kd}^\dagger \tilde{W} G_{kd}\right)\right|$$
$$= \frac{1}{N}\log\det\left(G_{1d}^\dagger \tilde{W} G_{1d}\right)$$
$$+ \frac{1}{N}\sum_{k\neq 1}\log\det\left(G_{kd}^\dagger \tilde{W} G_{kd}\right)$$
$$\leq \frac{1}{N}\log\left(\prod_i^n \hat{\boldsymbol{g}}_{1di}^\dagger \tilde{W} \hat{\boldsymbol{g}}_{1di}\right)$$
$$+ \frac{1}{N}\sum_{k\neq 1}\log\left(\prod_i^{n+1} \hat{\boldsymbol{g}}_{kdi}^\dagger \tilde{W} \hat{\boldsymbol{g}}_{kdi}\right) \quad (23)$$
$$< \frac{1}{N}\sum_i^n \hat{\boldsymbol{g}}_{1di}^\dagger \tilde{W} \hat{\boldsymbol{g}}_{1di}$$
$$+ \frac{1}{N}\sum_{k\neq 1}\sum_i^{n+1} \hat{\boldsymbol{g}}_{kdi}^\dagger \tilde{W} \hat{\boldsymbol{g}}_{kdi} \quad (24)$$
$$= \frac{1}{N}\sum_k \mathrm{Tr}\left[G_{kd}^\dagger \tilde{W} G_{kd}\right] \quad (25)$$

where (23) applies Hadamard's inequality [4] and (24) follows from $x > \log x$ for $x > 0$. Vector $\hat{\boldsymbol{g}}_{kdi}$ is the $i$th column of $G_{kd}$.

Next, we express the sum of traces in (25) as a trace of sums,
$$f_{ub}(\tilde{\boldsymbol{w}}) = \frac{1}{N}\mathrm{Tr}\left[\tilde{W}\sum_k G_{kd}^\dagger G_{kd}\right]$$

and substitute the interference alignment constraints (8) - (10) into (17), (19), and (21) to arrive at

$$f_{ub}(\tilde{\boldsymbol{w}}) = \frac{1}{N}\mathrm{Tr}\left[\tilde{W}\sum_{j\neq k} H_{jk}\Gamma_k\Gamma_k^\dagger H_{jk}^\dagger\right]$$
$$= \frac{1}{N}\sum_i \tilde{w}_i \sum_k c_{ki}\boldsymbol{\gamma}_{ki}\boldsymbol{\gamma}_{ki}^\dagger \quad (26)$$

where $\boldsymbol{\gamma}_{ki}$ is the $i$th row vector of matrix $\Gamma_k$ and
$$c_{ki} := \sum_{j\neq k}\left|(H_{jk})_{ii}\right|^2$$

for $k = \{1,2,3\}$, $j = \{1,2,3\}$, and $i = \{1,2,...,N\}$.

We then upper bound (26) by replacing the coefficients $c_{ki}$ with the single coefficient $c$ such that $c := \max\{c_{ki}\}$ for all $k$ and $i$. The final expression of the upper bound $f_{ub}(\tilde{\boldsymbol{w}})$ is then

$$f_{ub}(\tilde{\boldsymbol{w}}) = \frac{c}{N}\sum_i \tilde{w}_i \sum_k \boldsymbol{\gamma}_{ki}\boldsymbol{\gamma}_{ki}^\dagger = \frac{c}{N}\mathrm{Tr}\left[\tilde{W}\sum_k \Gamma_k\Gamma_k^\dagger\right].$$

Thus, the lower bound to the sum rate (15) is defined as

$$f_{lb}(\tilde{\boldsymbol{w}}) := \frac{3n+1}{2n+1}\log p + \frac{3}{N}\log\det \tilde{W}$$
$$+ \frac{1}{N}\sum_k \log\det\left(G_{kn}G_{kn}^\dagger\right)$$
$$- \frac{c}{N}\mathrm{Tr}\left[\tilde{W}\sum_k \Gamma_k\Gamma_k^\dagger\right]. \quad (27)$$

Note that the slope of the first term on the righthand side of the above states that the lower bound, as a function of $\log p$, increases linearly with a slope of $(3n+1)/(2n+1)$ as is the case with (15) when the limit of $p$ approaches infinity, thus maintaining the achievable degrees of freedom of the CJ scheme.

## IV. Suboptimal Improved Subspace Design

Based upon the lower bound to the sum rate (27), we present the main result of this paper summarized in the following theorem.

*Theorem 1:* The maximization of the sum rate lower bound defined in (27) for the $K = 3$ SISO frequency selective interference channel is defined as

$$C_{lb} = \max_{\tilde{\boldsymbol{w}}} f_{lb}(\tilde{\boldsymbol{w}})$$

subject to

$$\sum_{k=1}^3 \mathrm{Tr}\left[\Gamma_k^\dagger \tilde{W}\Gamma_k\right] = 3N, \; \tilde{w}_i > 0 \text{ for all } i \quad (28)$$

for which the closed-form solution is defined as

$$\tilde{w}_i := 3\left(\sum_k \boldsymbol{\gamma}_{ki}\boldsymbol{\gamma}_{ki}^\dagger\right)^{-1} \quad (29)$$

for all $k = \{1,2,3\}$ and $i = \{1,2,...,N\}$.

While an optimal solution to the maximization of the complete sum rate (15) is proven to exist in [3], it nonetheless must be computed using a constrained numerical optimization routine. Another suboptimal solution to the constrained sum rate maximization problem is also proposed in [3] that while less complex to solve than the optimal solution, is still non-trivial. However, the closed form solution to the maximization problem above while suboptimal to those proposed in [3] is far less complex to compute and still achieves significant gains in data rate performance as will be demonstrated in the upcoming section.

To derive the solution (29), we define the Lagrangian and gradient for the above constrained maximization problem respectively as

$$L\left(\tilde{\boldsymbol{w}}, \lambda\right) := f_{lb}\left(\tilde{\boldsymbol{w}}\right) - \lambda \left( \sum_k \text{Tr}\left[\Gamma_k^\dagger \tilde{W} \Gamma_k\right] - 3N \right)$$

and

$$\nabla_{\tilde{\boldsymbol{w}}} L\left(\tilde{\boldsymbol{w}}, \lambda\right) := \frac{3}{N}\text{diag}\left(\tilde{W}^{-1}\right) - \left(\frac{c}{N} + \lambda\right) \text{diag}\left(\sum_k \Gamma_k \Gamma_k^\dagger\right).$$

Next, using the constraint (28), we set the gradient above to zero and multiply it by $\tilde{\boldsymbol{w}}^t$,

$$\tilde{\boldsymbol{w}}^t \ \nabla_{\tilde{\boldsymbol{w}}} L\left(\tilde{\boldsymbol{w}}, \lambda\right) = 3 - \left(\frac{c}{N} + \lambda\right) 3N = 0.$$

Solving for $\lambda$, we get $\lambda = (1-c)/N$. Then setting the gradient to zero, substituting in the above solution for $\lambda$, and solving for $\tilde{\boldsymbol{w}}$, we arrive at (29).

## V. SIMULATION RESULTS

In this section, we provide numerical results to compare the data rate performance of the proposed design, those presented in [3], and the original CJ scheme.[1] Data is presented for the following algorithms abbreviated as

- CJ - the original CJ scheme
- $\text{KT}_\text{op}$ - the optimal design proposed by KT in [3]
- $\text{KT}_\text{sop1}$ - the first suboptimal design proposed by KT in [3]
- $\text{KT}_\text{sop2}$ - the suboptimal design proposed herein.

Also provided as a reference is the line $D_N \log_2(\text{snr})$ where $D_N = \{\frac{4}{3}, \frac{16}{11}\}$ for $N = \{3, 11\}$, respectively.

It was shown in [3] that while significant gains in sum rate were achievable using the proposed algorithms therein, even further gains were achievable when combined with SHV orthonormalization [2]. Therefore, the simulation results presented for the schemes listed above, with the exception of CJ, have been obtained in combination with SHV orthonormalization. Summaries of $\text{KT}_\text{op}$, $\text{KT}_\text{sop1}$, and SHV orthonormalization are available in the Appendix.

As shown in Fig. 1 for $N = 3$, the optimal $\text{KT}_\text{op}$ design achieves a gain in sum rate of 2.39 bits/s/Hz over the original CJ scheme in the high SNR regime at 50dB. Just beneath that, the first suboptimal design, $\text{KT}_\text{sop1}$, is only 0.22 bits/s/Hz less than the optimal design; and the proposed suboptimal closed-form solution, $\text{KT}_\text{sop2}$, is only 0.47 bits/s/Hz less than the optimal design.

[1]Channel coefficients are drawn i.i.d. from a complex Gaussian distribution. However, the channel coefficients were further bounded to be within a non-zero minimum value and a finite maximum value in order to avoid degenerate channel conditions.

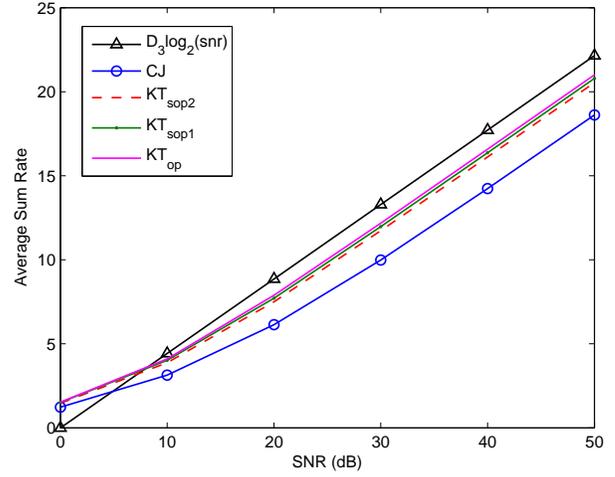

Fig. 1. Average sum rate comparison under various schemes for N = 3.

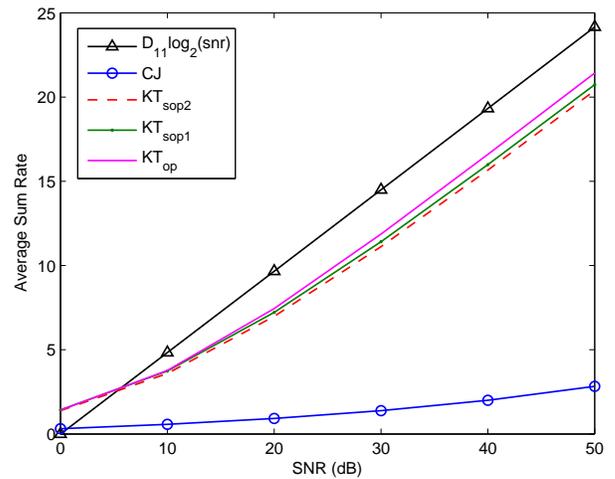

Fig. 2. Average sum rate comparison under various schemes for N = 11.

In the mid SNR regime at 10dB, the difference between $\text{KT}_\text{op}$ and $\text{KT}_\text{sop2}$ is 0.23 bits/s/Hz, approximately half the difference of that at 50dB. Moreover, a 0.73 bits/s/Hz improvement in sum rate over the CJ scheme is achievable with use of the simpler $\text{KT}_\text{sop2}$ design.

Similar improvements are achievable for larger symbol extensions of the channel. As shown in Fig. 2 for $N = 11$ at an SNR of 50dB, the difference between $\text{KT}_\text{op}$ and $\text{KT}_\text{sop2}$ is 1.04 bits/s/Hz while the difference between $\text{KT}_\text{sop2}$ and the CJ scheme is 17.56 bits/s/Hz. At a mid SNR of 10dB, the difference between $\text{KT}_\text{op}$ and $\text{KT}_\text{sop2}$ is 0.18 bits/s/Hz while the difference between $\text{KT}_\text{sop2}$ and the CJ scheme is 3.02 bits/s/Hz.

## VI. CONCLUSION

In this paper, we have presented the design of a new low complexity algorithm that improves the sum rate performance of the Cadambe and Jafar [1] $K = 3$ user interference alignment scheme while preserving its achievable degrees of freedom. The proposed design is based upon the subspace optimization algorithm introduced by Kim and Torlak in [3] which maximizes the network sum rate with respect to $w$. Unlike the optimal solution in [3], the proposed design does not require use of a numerical optimization solver. Rather, the proposed design has a closed-form solution that, while suboptimal, still achieves significant gains in data rate performance at far less of a computational cost. The proposed algorithm can be extended to a greater number of users and to other variations of the original CJ scheme [5] [6].

## APPENDIX A
### SUMMARY OF KT IMPROVED SUBSPACE OPTIMIZATION

In [3], the sum rate (15) for the $K = 3$ user SISO Gaussian constant interference channel is proven to be a concave function with respect to $\tilde{w}$ that can be expressed in the following optimization problem as

$$C = \min_{\boldsymbol{q}:q_i \leq c_i} \min_{\lambda > 0} \max_{\tilde{\boldsymbol{w}} \succ 0} \left[ f(\tilde{\boldsymbol{w}}) - \lambda \left( \boldsymbol{q}^t \tilde{\boldsymbol{w}} - 3N \right) \right] \quad (30)$$

where the elements of the $N$ by 1 vectors $\boldsymbol{q}$ and $\boldsymbol{c}$ are defined respectively as

$$q_i := c_i - \lambda_i/\lambda$$
$$c_i := \sum_k \|\boldsymbol{\gamma}_{ki}\|^2$$

and $\boldsymbol{\gamma}_{ki}$ is the $i$th row of $\Gamma_k$. Thus there exists a maximum $C$ for some $\tilde{\boldsymbol{w}} = \boldsymbol{d}^*$ which satisfies

$$\frac{1}{c_i} \sum_k (A_{ki} - B_{ki}) = \lambda^*, \text{ if } \tilde{w}_i > 0$$
$$\frac{1}{c_i} \sum_k (A_{ki} - B_{ki}) \leq \lambda^*, \text{ if } \tilde{w}_i = 0$$

for some $\lambda^* > 0$. Terms $A_{ki}$ and $B_{ki}$ are defined as

$$A_{ki} := \frac{p}{N} \boldsymbol{g}_{kni} \left( I + p G_{kn}^\dagger \tilde{W} G_{kn} \right)^{-1} \boldsymbol{g}_{kni}^\dagger$$
$$B_{ki} := \frac{p}{N} \boldsymbol{g}_{kdi} \left( I + p G_{kd}^\dagger \tilde{W} G_{kd} \right)^{-1} \boldsymbol{g}_{kdi}^\dagger$$

where $\boldsymbol{g}_{kni}$ and $\boldsymbol{g}_{kdi}$ are the $i$th row vectors of matrices $G_{kn}$ and $G_{kd}$, respectively.

## APPENDIX B
### SUMMARY OF KT SUBOPTIMAL SUBSPACE CONDITIONING

In [3], a suboptimal maximization problem to that of (30) is defined as

$$C_{lb} = \max_{\tilde{\boldsymbol{w}}} f_{lb}(\tilde{\boldsymbol{w}})$$

subject to

$$\sum_{k=1}^3 \text{Tr}\left[ \Gamma_k^\dagger \tilde{W} \Gamma_k \right] = 3N, \ \tilde{w}_i > 0 \text{ for all } i \quad (31)$$

where a lower bound to the sum rate (15) is defined asymptotically in SNR as

$$f_{lb}(\tilde{\boldsymbol{w}}) := \frac{3n+1}{N} \log p + \frac{3}{N} \log \det \tilde{W}$$
$$+ \frac{1}{N} \sum_k \log \det \left( G_{kn} G_{kn}^\dagger \right) - \frac{1}{N} \text{Tr}\left[ \tilde{W} \sum_k G_{kd} G_{kd}^\dagger \right]. \quad (32)$$

The solution to the above maximization problem is

$$\tilde{w}_i = \frac{3}{N} \left( \lambda_{lb} \sum_k a_{ki} + \sum_k b_{ki} a_{ki} \right)^{-1} \quad (33)$$

where

$$a_{ki} := \gamma_{ki} \gamma_{ki}^\dagger \quad (34)$$
$$b_{ki} := \frac{1}{N} \sum_{j \neq k} |(H_{jk})_{ii}|^2 \quad (35)$$

for $j = \{1, 2, 3\}$ and $\lambda_{lb}$ must satisfy

$$\lambda_{lb} > -\frac{\sum_k b_{ki} a_{ki}}{\sum_k a_{ki}}$$

for all $i$.

## APPENDIX C
### SUMMARY OF SHV ORTHONORMALIZATION

In [2], for the $K = 3$ SISO interference alignment channel, Shen, Host-Madsen, and Vidal show that the high SNR offset of the sum rate, eq. (5) in [2], can be maximized under constraints (8) - (20) in [2] by precoding the CJ transmit beamforming vectors $V_2$ and $V_3$ by the $n$ by $n$ matrices $F$ and $E$, respectively, such that

$$\tilde{V}_2 := V_2 F$$
$$\tilde{V}_3 := V_3 E$$

and matrix $V_1$ is unaltered, defined as in the original CJ scheme.

They show that the solution to $F$ and $E$ orthogonalizes the $n$ columns of $V_2$ and $V_3$, respectively, while simultaneously allocating $N/n$ units of power to each column. SHV also show that by doing so, the solutions uphold the interference alignment constraints (b) in [2], maintaining the achievable degrees of freedom proposed by CJ.